# How does the substrate affect the Raman and excited state spectra of a carbon nanotube?

Mathias Steiner[1], Marcus Freitag[1], James C. Tsang[1], Vasili Perebeinos[1], Ageeth A. Bol[1], Antonio V. Failla[2], and Phaedon Avouris[1*]

[*]_avouris@us.ibm.com_

[1]_IBM Thomas J. Watson Research Center, Yorktown Heights, New York 10598, USA_

[2]_Imaging and Microscopy Laboratory, Cambridge Research Institute of Cancer Research UK, Cambridge CB2 0RE, UK_

**Abstract:** We study the optical properties of a single, semiconducting single-walled carbon nanotube (CNT) that is partially suspended across a trench and partially supported by a SiO$_2$-substrate. By tuning the laser excitation energy across the $E_{33}$ excitonic resonance of the suspended CNT segment, the scattering intensities of the principal Raman transitions, the radial breathing mode (RBM), the D mode and the G mode show strong resonance enhancement of up to three orders of magnitude. In the supported part of the CNT, despite a loss of Raman scattering intensity of up to two orders of magnitude, we recover the $E_{33}$ excitonic resonance suffering a substrate-induced red shift of 50 meV. The peak intensity ratio between G band and D band is highly sensitive to the presence of the substrate and varies by one order of magnitude, demonstrating the much higher defect density in the supported CNT segments. By comparing the $E_{33}$ resonance spectra measured by Raman excitation spectroscopy and photoluminescence (PL) excitation spectroscopy in the suspended CNT segment, we observe that the peak energy in the PL excitation spectrum is red-shifted by 40 meV. This shift is associated with the energy difference between the localized exciton dominating the PL excitation spectrum and the free exciton giving rise to the Raman excitation spectrum. High-resolution Raman spectra reveal substrate-induced symmetry breaking, as evidenced by the appearance of additional peaks in the strongly broadened Raman G band. Laser-induced line shifts of RBM and G band measured on the suspended CNT segment are both linear as a function of the laser excitation power. Stokes/anti-Stokes measurements, however, reveal an increase of the G phonon population while the RBM phonon population is rather independent of the laser excitation power.

## 1. Introduction

The optical and electronic properties of semiconducting single-walled carbon nanotubes (CNTs), quasi one-dimensional systems on the nanometer scale with a diameter-related band-gap, are promising for a number of technological applications [1]. However, it has been recognized that contact with a (dielectric) substrate can significantly modify their favorable optical emission and electrical transport characteristics and, hence, affect the performance of CNT-based optoelectronic devices [2, 3]. Raman and photoluminescence (PL) spectroscopies have turned out to be the most valuable tools for measuring electron phonon coupling in CNTs as well as the coupling of CNTs to their local environment, see e.g. [4-8]. However, the experimental difficulties in performing excitation spectroscopy on the single tube level in general, as well as the suppression of emission when CNTs are brought in contact with a substrate, have prevented systematic experimental studies until recently, even though substrate-induced changes in the Raman properties of individual CNTs have been known (see e.g. [9]).

In this article, we investigate how the contact with a dielectric ($SiO_2$)-substrate modifies the optical properties of the same spatially isolated CNT. Using resonance Raman excitation spectroscopy we map the $E_{33}$ excitonic resonance, both in the freely suspended segment of a spatially isolated CNT and in a part of the same CNT that is supported by a $SiO_2$-substrate. We discuss the differences in spectral peak positions and widths of the $E_{33}$ resonance on the suspended CNT segment measured with Raman excitation spectroscopy and PL excitation spectroscopy in the context of exciton localization. We identify the different interactions that lead to the strong differences in the Raman scattering intensities measured on the suspended and supported segments of the same CNT. We analyze the shifts and broadenings of individual Raman transition bands in the context of symmetry breaking due to the presence of the $SiO_2$-substrate. Finally, we investigate the line shifts and the anti-Stokes/Stokes intensity ratios of the RBM and the G modes in the two segments induced by the laser itself.

## 2. Experimental section

We measured Raman spectra using a scanning optical microscope equipped with a standard microscope objective (100x, NA=0.8, Nikon) providing a focal spot diameter of about 500 nm and a excitation power density of up to 180 kW/cm$^2$. Despite for the investigation of laser-induced line shifts and heating, we chose excitation power densities low enough to avoid spectral shifts and broadenings. A feedback-controlled piezoelectric scanning stage (P-527.2CL, PI) accomplished raster- scanning of the CNT with respect to the microscope objective with nanometer-precision. As excitation light sources, we used an Ar$^+$ laser (Innova 300, Coherent) operated between 476.5 nm and 514.5 nm, a frequency-doubled Nd:YVO4 laser (Millenia, Spectra-Physics) at 532 nm, a tunable dye laser (CR-599 operated with R6G solution, Coherent), as well as a HeNe laser (1144 P, JDS Uniphase) at 632.8 nm. The Raman-scattered light was separated from the laser excitation light using holographic notch filters (Kaiser) and spectrally analyzed using either (A) a spectrograph (Triax 322, Jobin Yvon/Horiba) equipped with gratings having a groove density of 300 mm$^{-1}$ and 1200 mm$^{-1}$ and a LN$_2$-cooled charge coupled device (Spectrum One, Jobin Yvon/Horiba) or (B) a spectrographic stage (of the triple-stage spectrograph XY, Dilor) equipped with a grating having a groove density of 1200 mm$^{-1}$ and a LN$_2$-cooled and IR-enhanced deep-depleted charge coupled device (Jobin Yvon/Horiba). With (A), we achieve a spectral resolution of up to 8cm$^{-1}$ while obtaining 1.8cm$^{-1}$ with configuration (B). We chose acquisition times ranging between 30 seconds and 300 seconds to achieve sufficient signal-to-noise-ratios in the measured Raman spectra. A corresponding background spectrum measured with the same set of acquisition parameters was subtracted from each measured Raman spectrum to improve data quality. PL (excitation) spectra were acquired with the same microscopy setup, extended with a liquid-nitrogen cooled HgCdTe-detector (IRLabs) and a transmission grating fabricated on top of a prism delivering a spectral resolution of 25nm. We corrected the measured PL spectra for the wavelength-dependent detection efficiency of the experimental setup. For all measurements reported in this paper, we aligned the long axis of the CNT parallel to the polarization axis of the excitation laser.

### 3. Results and discussion

#### 3.1 *Sample characterization*

Single-walled carbon nanotubes were grown across micron-wide trenches in a Si/SiO$_2$-substrate by chemical vapor deposition (CVD) using a method similar to that reported in [10]. As shown in Fig. 1(a), we obtain spatially isolated, individual CNTs that are partially supported by the SiO$_2$-substrate and partially suspended over the trench. The scanning electron microscopy and Raman microscopy images shown in Fig. 1(a) demonstrate that we can spatially resolve the suspended and supported parts of the CNT. A Raman spectrum measured on the suspended CNT segment is shown in Fig. 1(b). The spectrum is dominated by transitions associated with the radial breathing (RBM) and the main tangential C-C stretching (G) phonon modes [4, 5]. Moreover, we observe a transition due to the Raman forbidden but defect-induced D mode, attesting to the presence of defects. The frequency of the RBM (142 cm$^{-1}$ in the present case) is inversely proportional to the diameter $d_t$ of the CNT and by using the relations $d_t = 248/\Omega_{RBM}$ [11] and $d_t = 227/\Omega_{RBM}$ [12] we obtain $d_t$ = 1.7 nm and $d_t$ = 1.6 mm, respectively. The PL emission in the near infrared spectral range (see spectrum Fig. 1(c)) is associated with the radiative decay of the $E_{11}$-excitons ($E_{11}$ = 0.616 eV in the present case) and confirms the semiconducting character of the CNT studied [13-15]. Since the enhancement of the excitation of the phonon modes is spectrally localized roughly in the range between 490 nm and 630 nm (corresponding to an energy range between 2 eV and 2.5 eV), we conclude that the measured Raman spectrum in Fig. 1(b) is resonantly enhanced by the $E_{33}$-resonance at about 2.2 eV (see, for example, the $E_{ii}$-$d_t$-relation (Kataura-plot) in [4]).

From the Raman scan image in Fig. 1(a), we recognize that the Raman scattering intensities measured on suspended and supported parts of the same CNT vary significantly. In the following, we systematically study these variations by analyzing the integrated intensities of the RBM, D and G modes as a function of the laser excitation energy, i.e. using the resonance-Raman excitation spectra.

## 3.2 Raman excitation spectra of suspended and supported CNT segments

In Fig. 2(a), we plot separately the integrated scattering intensities of the Raman transition bands RBM, D and G measured on the suspended CNT segment as functions of the laser excitation energy. Both RBM and G band intensities vary strongly as a function of the laser excitation energy, spanning an intensity range of three orders of magnitude due to the electronic resonance enhancement induced by the $E_{33}$-level. The maximum scattering intensities for both the RBM and G mode coincide at 2.210±0.006 eV, the peak energy of the $E_{33}$-level of the CNT in the suspended part, and we estimate a resonance width of 50±20 meV. As expected, we cannot resolve the splitting of 17 meV associated with the resonances of the incoming and the scattered photons in the case of the low-energy RBM, while we clearly observe both ingoing and outgoing resonances for the high-energy G mode, which are spectrally separated by about 190 meV. However, both the peak energy and line shape of the D band Raman excitation spectrum differ significantly from those of the RBM and G modes. Moreover, while the peak intensity ratios between suspended and supported CNT segments are 23 for the G band and 38 for the RBM, respectively, we observe only moderate intensity variations for the D band by up to a factor of 2, as obtained by comparison with the spectra shown in Fig. 2(b). Note that, due to the shift of the $E_{33}$ resonance, the intensity ratio at a specific laser excitation energy can be significantly larger (15 for the D band, 56 for the G band, 116 for the RBM).

    In the segment of the same CNT that is supported by the SiO$_2$-substrate (see Fig. 2(b)), we observe that the scattering intensities of the RBM and G band are reduced by up to two orders of magnitude while the $E_{33}$ resonance is red shifted by about 55±10 meV and has a width of 75±10 meV (i.e. broadened by a factor of 1.5 as compared to the suspended CNT segment). In quasi-one-dimensional CNTs, the local environment is largely responsible for the electron-hole screening and, as a result, the properties of the bound electron-hole pairs (excitons) depend strongly on the dielectric environment [3]. While the lowest exciton binding energy scales with the dielectric constant as $\varepsilon^{-1.4}$ [16], the band gap renormalization is inversely proportional to the dielectric environment $\varepsilon^{-1.0}$ [17]. Therefore, we expect the exciton excitation energy (bandgap minus exciton binding energy) to decrease with increasing dielectric screening, consistent with our experimental observations. However, the energy shift of the $E_{22}$-exciton in a suspended CNT induced by a change in the local dielectric environment has recently been reported to be only around 30 meV [18]. This implies that an energy shift induced by the dielectric (SiO$_2$)-substrate (only present in one half-space) cannot account alone for the large red shift observed in our experiments. We attribute the energy shift observed as mainly due to the energy difference between free and localized excitons as will be discussed in more detail in the following subsection.

    As in the case of the suspended part of the CNT, the D band intensity characteristics differ from those measured for RBM and G as we vary the laser excitation energy. To illustrate these differences, we plot the ratio between the integrated G band intensity and the integrated D band intensity as a function of the laser excitation energy in Fig. 2(c). The intensity ratio G/D for specific laser excitation energies varies between 3 and 70 in the suspended part and between 2 and 20 in the supported part of the CNT while the G/D peak intensity ratio is 39 in the suspended CNT segment and 3.5 in the

supported CNT segment. By assuming that the ratio between the measured peak intensities of both G and D is inversely proportional to the scattering center ("defect") density [4] that give rise to the D-band, we find that the defect density in the supported CNT segment is increased by a factor of about 10.

### 3.3 *CNT excitation spectra and exciton localization*

In Fig. 3, we compare the $E_{33}$ resonance excitation spectra that have been measured on the suspended CNT segment by integrating independently the RBM Stokes intensity and the $E_{11}$-PL intensity as a function of the laser excitation energy. The PL excitation spectrum has its spectral peak position at 2.174±0.005 eV and a spectral width of 120±20 meV, which is much broader than the $E_{22}$-widths of around 40 meV that have been measured for individual CNTs with a comparable method [19]. As compared to the PL excitation spectrum, the width of the Raman excitation spectrum measured on the same CNT is significantly smaller, 30±2 meV, and the intensity maximum at 2.216±0.002 eV is blue shifted by around 40 meV. It is generally assumed that the two different spectroscopies would give the same result. However, the spectral shifts and width observed here suggest that they may probe the same state but in two different environments. There is increasing evidence that the optics of CNTs involves localized excitons and that localized excitons are formed at defect sites, see e.g. [20]. It is then reasonable to assume that while the resonance-Raman excitation spectrum reflects the absorption of the extended CNT, i.e. free exciton states, the PL excitation spectrum is biased towards the absorption at defect sites that trap the excitons. The red shift of the PL excitation spectrum by 40 meV is then a reflection of the lower excitation energy at the defect sites (see Fig. 4), while the larger width reflects both the heterogeneity of these sites and their typically stronger electron-phonon coupling [21]. In the following, we will discuss qualitatively the measured widths of both excitation spectra in the context of this interpretation.

The resonance-Raman scattering cross section $R_0$ of the illuminated CNT segment is determined by the dipole matrix elements between the ground state $|GS\rangle$ and the resonant excited state $|\Psi_{n,ii}\rangle$:

$$R_0 \propto \left| \sum_{n=1}^{N} \frac{\langle GS|\hat{p}|\Psi_{n,ii}\rangle \langle \Psi_{n,ii}|H_{exc-ph}|\Psi_{n,ii}+phonon\rangle \langle \Psi_{n,ii}+phonon|\hat{p}|GS+phonon\rangle}{(E_L - E_{ii} + i\gamma_{ii})(E_L - E_{ii} - \hbar\Omega_{ph} + i\gamma_{ii})} \right|^2 \quad (1)$$

where $E_L$ is the laser excitation energy, $\gamma_{ii}$ is the width (decay rate) of the resonant exciton-state $|\Psi_{n,ii}\rangle$ with energy $E_{ii}$, $H_{exc-ph}$ is the exciton-phonon Hamiltonian, and $\hbar\Omega_{ph}$ is the phonon energy. From Eq. (1), we expect that the width in the resonance-Raman excitation spectrum is reduced by a factor of $(\sqrt{2}-1)^{0.5} \approx 0.64$ as compared to the corresponding PL excitation spectrum, at least for small phonon energies. However, if we scale the measured width of the PL excitation spectrum in order to compare it with the measured width of the resonance-Raman excitation spectrum, we find $\gamma_{33}^{PLE} = 0.64 \cdot 120 \text{meV} = 77 \text{meV} \neq 30 \text{meV} = \gamma_{33}^{RRS}$. In terms of our previous discussion this implies that the width of the $E_{33}$-level as probed by absorption through localized

excitons (i.e. in the PL excitation spectrum) is broadened by a factor of $\gamma_{33}^{PLE}/\gamma_{33}^{RRS} = 2.6$ compared to the width of the $E_{33}$-level as probed by absorption through free excitons (i.e. in the Raman excitation spectrum). From this result and by assuming that the intrinsic width of high-energy excitons is dominated by the decay rate of excitons into optical phonons [22, 23], we estimate that the ratio between the exciton-phonon couplings $M_{33} = \langle \Psi_{33} | H_{exc-ph} | \Psi_{33} + phonon \rangle$ of free and localized excitons is of the order of $M_{33}^{localized}/M_{33}^{free} = (\gamma_{33}^{PLE}/\gamma_{33}^{RRS})^{0.5} = 1.6$.

Consistent with previous studies, we do not observe PL emission from the supported segments of the CNT. The non-radiative CNT excited state decay rate is increased due to the additional relaxation pathways provided by the substrate. As a result, the PL emission in the supported CNT segment drops below the detection limit of our experimental setup.

### 3.4 *Raman scattering intensities of suspended and supported CNT segments*

We now discuss three substrate-induced effects that are responsible for the measured Raman intensity variations in supported and suspended segments of the CNT. Specifically: (A) the broadening of the *E₃₃*-resonance, (B) the loss of oscillator strength due to exciton localization and (C) the modification of the intensity distributions of the focused laser excitation field.

A 50%-increase of the full-width-at-half-maximum-value (FWHM) of the peak width in the Raman excitation spectrum in the supported CNT segment as observed in our experiments corresponds to a broadening of the electronic level by a factor of 1.5. As a result of the broadening effect (A), the peak intensity in the supported CNT segment is reduced by a factor $1.5^3 = 3.4$ (see Eq. (1)).

In order to estimate contribution (B), we assume that the oscillator strength $f_{ii} \propto |\langle GS | \hat{p} | \Psi_{n,ii} \rangle|^2$ in Eq. (1) is inversely proportional to the defect density in the illuminated CNT segment. This assumption is motivated by our experimental observations on different CNT samples. In our analysis of the measured G/D-Raman intensity ratios in both supported and suspended CNT segments (see Fig.2(c)), we concluded that the defect density in the CNT is increased by a factor of about 10 in the presence of the substrate. Therefore, we estimate, very roughly, that a 10-fold increase in defects (or, in other words, trapping sites for exciton localization) in the supported CNT segment will translate into an about 10-fold decrease in Raman scattering intensity as compared to the suspended CNT segment.

Finally, we estimate the contribution (C) that originates from modifications of the intensity distribution of the focused laser excitation field at the position of the two CNT segments. In order to model the intensity distributions we decompose the incident laser beam, a linearly polarized Hermite-Gaussian TEM$_{00}$-mode, into elementary plane waves, and propagate each individual component through the focusing optics by accounting for the boundary conditions at the occurring interfaces using the angular spectrum representation [24]:

$$\mathbf{E}(x,y,z)_f = \frac{ife^{-ikf}}{2\pi k_z} \int_0^{k_{x,\max}} \int_0^{k_{y,\max}} \mathbf{E}_\infty(k_x, k_y) e^{-i(k_x x + k_y y + k_z z)} dk_x dk_y \qquad (2)$$

Here, f is the focal length of the microscope objective and $k = |\mathbf{k}| = \sqrt{k_x^2 + k_y^2 + k_z^2}$ is the wave vector associated with each individual plane wave satisfying $0 \leq k \leq f \sin\theta_{\max}$. The maximum focusing angle $\theta_{\max}$ is given by the numerical aperture $NA = n_{air} \sin\theta_{\max}$ of the microscope objective ($NA = 0.8$ in the present case) and the field distribution of the collimated laser beam at the surface of the microscope objective $\mathbf{E}_\infty(k_x, k_y)$ is determined by the boundaries of both incident laser beam and focusing optics [24]. We include the existing optical interfaces and derive the associated reflections and transmissions based on the dielectric functions of the constituents (air, $SiO_2$, Si).

In our model, we consider the actual sample geometry sketched in Fig. 1(a). We define the ($x,y,z=0$)-plane as the plane of the air/$SiO_2$-interface, or, from another point of view, the plane of the CNT. The incident laser beam is polarized in x-direction (in our experiments, the CNT is aligned along the polarization axis of the light). We now consider two cases (A) and (B). In case (A), the incident laser field $\mathbf{E}_f(x,y,z)$ has its focus at position $z=0$ in air, 500nm above the Si-reflector. In our experiments, this corresponds to the position of the suspended CNT segment. To improve the modeling accuracy, we include the field contribution $\mathbf{E}_{r,Si}(x,y,z)$ that has been back-reflected by the Si-reflector into the focal plane at $z=0$:

$$\mathbf{E}_{(A)}(x,y,z) = \mathbf{E}_f + \mathbf{E}_{r,Si} \qquad (3)$$

The calculated intensity cross section is shown in Fig. 5(a). For the laser wavelength used ($\lambda_{laser} = 567$ nm), the intensity maximum of the focal field distribution occurs close to the ($z=0$)-position, i.e. the position of the CNT.

In case (B), we include the $SiO_2$-substrate and sum up the field contributions that arise from the transmissions and reflections at the respective interfaces in the following way:

$$\mathbf{E}_{(B)}(x,y,z) = \mathbf{E}_f + \mathbf{E}_{r,SiO_2} + \mathbf{E}_{r,Si} \text{ if } z \leq f \text{ and}$$
$$\mathbf{E}_{(B)}(x,y,z) = \mathbf{E}_{t,SiO_2} + \mathbf{E}_{r,Si} \text{ if } z > f \qquad (4)$$

The resulting intensity cross section is shown in Fig. 5(b). While the intensity maximum of the focal field distribution still occurs close to the ($z=0$)-position, the intensity distribution is modified significantly by the presence of the $SiO_2$-substrate.

By comparing the intensities for each individual component at the ($x=0,y=0,z=0$)-position (see Fig. 5(c)), we find that the intensity contribution associated with the polarization axis is the strongest, as expected, and decreases with a smooth slope as the wavelength increases. The intensities associated with the $y$ ($z$) component of the laser field are three (one) orders of magnitude weaker. In Fig. 5(d), we plot the amplification factors that have been obtained by normalizing the focal intensities in air (case (A), corresponding to the suspended CNT segment) with the focal intensities on the air/SiO2-interface (case (B), corresponding to the supported CNT segment). As a result, the suppression of the laser excitation intensity at the position of the supported CNT segment varies as a function of the wavelength and reaches factors of up to 2.5.

We can now estimate an upper limit of the overall reduction of the Raman peak intensity in the supported CNT segment by multiplying the contributions from (A), (B) and (C) obtained above and thus get 3.4 x 10 x 2.5 = 85. This value overestimates the peak intensity reduction observed for the RBM (38) by a factor of 2.2 and G band (23) by a factor of 3.6. We attribute the discrepancy primarily to our rough estimate of contribution (B). We also speculate that the measured differences in the peak intensities of RBM and G modes are likely due to a different scaling of the exciton-phonon matrix elements for these modes upon exciton localization.

### 3.5 *Substrate-induced modifications of CNT Raman transitions*

In the following, we analyze the effect of the $SiO_2$-substrate on the center frequency, line width and line shape of the individual Raman transition bands associated with RBM, D and G phonon modes using high-resolution Raman spectroscopy. In Fig. 6, we compare for each phonon mode the Raman bands that have been measured on the suspended, or the supported part of the same CNT. For the RBM (center: 142 cm$^{-1}$, width: 10 cm$^{-1}$), we observe only a very small substrate-induced increase in both center frequency and line width in the range of 1cm$^{-1}$, depending on the position on the substrate, in agreement with other recent results [9, 25]. The D band measured on the supported part clearly exhibits a frequency upshift of 4 cm$^{-1}$ and a broadening of 3 cm$^{-1}$ as compared to the spectrum measured on the suspended part (center: 1337 cm$^{-1}$, width: 18 cm$^{-1}$). For the G mode, we observe the strongest substrate-induced spectral effects. The spectrum on the suspended segment of the CNT exhibits a very strong and narrow $G^+$ contribution (center: 1590cm$^{-1}$, width: 5cm$^{-1}$) and a weak $G^-$ contribution spectrally centered around 1575 cm$^{-1}$ (see e.g. [9, 25, 26] for comparison). On the supported CNT segment, we observe that the strongest spectral component is up-shifted by 4 cm$^{-1}$ and broadened by a factor of 2.5, as compared to the original $G^+$ contribution, while we observe additional spectral components at the low energy side of the spectrum.

We correlate the appearance of the multiple G band components in the CNT Raman spectrum on the $SiO_2$-substrate with axial symmetry breaking. Here, the symmetry is broken by perpendicular fields induced by the substrate interaction, giving rise to similar spectral effects as observed by changing the polarization of the laser excitation light [27, 28]. As a result, the $E_{33}$-state is mixed with finite angular momentum states $E_{\Delta m=\pm 1,\pm 2,\pm 3,...}$. For example, states with $\Delta m = \pm 1$ can be excited by light polarized perpendicular to the CNT long axis. The amount of mixing can be quantified using a phenomenological parameter $\alpha_{\Delta m}$ that depends on the nature of the interaction and the energy splitting between the $E_{33}$ and the $E_{\Delta m}$ states. It is expected that $\alpha_{\Delta m}$ would decrease as $|\Delta m|$ increases. The wave function $|\Psi_{33}^{substrate}\rangle$ acquires an admixture of the finite angular momentum states $|\Psi_{\Delta m \neq 0}\rangle$:

$$|\Psi_{33}^{substrate}\rangle \propto |\Psi_{33,\Delta m=0}\rangle + \sum_{\Delta m \neq 0} \alpha_{\Delta m} |\Psi_{\Delta m}\rangle \qquad (5)$$

In principle, this mixing can also change the energy of the $E_{33}$ resonance. However, the sign and the magnitude of the shift depends not only on the coupling strength, but also on the energy splitting $E_{33} - E_{\Delta m}$, which makes it difficult to predict without knowing the detailed exciton band structure. Therefore, we constrain the discussion on the appearance of the finite angular momentum phonons in the measured Raman spectrum by considering only the first term of the sum in Eq. (5):

$$I \propto R\left[\left|M_{\Delta m=0}\right|^2 \delta(\omega - \omega_0) + 8\alpha^2_{\Delta m=\pm 1}\left|M_{\Delta m\pm 1}\right|^2 \delta(\omega - \omega_{\Delta m=\pm 1})\right] \quad (6)$$

where $R$ is the resonant part of the Raman scattering cross-section (see Eq. (1)) and the electron-phonon matrix element $M_{\Delta m}$ depends on the azimuthal angular momentum of the phonon $\omega_{\Delta m}$. Remarkably, we find that the measured peak energies labeled as A, $E_1$ and $E_2$ in the G-band spectrum measured on the supported CNT segment (see Fig. 6, lower panel) indeed correspond to the reported energies of the finite angular momentum G phonons [4]. The intensity ratio between the A phonon peaks ($\Delta m = 0$) is influenced by the local distortions induced by the substrate. By assuming that the exciton-phonon matrix elements for states with different angular momentum are equal, $\left|M_{\Delta m=0}\right| = \left|M_{\Delta m=\pm 1}\right|$, we can estimate $\alpha_{\Delta m=\pm 1} = 0.14$ by accounting for the measured intensity ratio between the $E_1$ peak and the (stronger) A peak.

### 3.6 *Laser-induced shifts, broadenings and heating of phonon modes in suspended and supported CNT segments*

We now compare the dependence of the RBM and the G band spectra of the suspended and supported segments of the same CNT on the laser excitation power. In the upper panels of Fig. 7 we plot a series of RBM and G band spectra that have been acquired on the suspended CNT segment for various laser excitation power densities. For both Raman bands, we observe reproducible downshifts of the center frequency in the range of 2 cm$^{-1}$ for laser excitation power densities as high as 150 kW/cm$^2$ and the best linear fits (see lower panels of Fig. 7) deliver $\Delta_{RBM}$= -0.014±0.001 cm$^{-1}$/kWcm$^{-2}$ and $\Delta_G$= -0.011±0.001 cm$^{-1}$/kWcm$^{-2}$. On the part of the CNT that is supported by the SiO$_2$-substrate, we do not observe significant variations of the center frequencies for either RBM or G bands when we increase the laser excitation power density up to 180 kW/cm$^2$ (see middle and lower panels of Fig. 7), indicating that laser irradiation affects the suspended and supported CNT segments differently.

In order to clarify the origin of the laser-induced frequency downshift of the G and RBM modes, we performed Stokes/anti-Stokes Raman spectroscopy. In Fig. 8, we observe similar laser-induced frequency shifts for both anti-Stokes and Stokes bands as we increase the laser excitation power density. Lorentzian fits to the experimental spectra provide estimates for the integrated intensities of each individual Raman band. The anti-Stokes/Stokes intensity ratio for the RBM remains almost identical, 0.830 and 0.829, as we increase the excitation power density from 15 kW/cm$^2$ to 150 kW/cm$^2$, indicating that the CNT remains at essentially at the same temperature. By contrast, the anti-Stokes/ Stokes intensity ratio for the G band increases by a factor of three, from 0.002 to 0.006, when we double the laser excitation power density from 75 kW/cm$^2$ to 150 kW/cm$^2$, indicating heating of this CNT phonon mode. Temperature variations have been shown to shift the electronic energy levels of CNTs [29]. Similarly, we found that the electronic resonance conditions in a CNT can be substantially modified upon self-heating induced by an electrical current, leading to shifts and broadening of the $E_{33}$-resonance [30]. However, we do not expect significant changes of the resonance conditions for moderate temperature changes in the range of 100 K. We estimate the G phonon temperature change $\Delta T_G$ for a laser power increase from 75 kW/cm$^2$ to 150 kW/cm$^2$ using the relation

$$\frac{I_{G,anti-Stokes}}{I_{G,Stokes}} = \exp\left\{\frac{\hbar\Omega_G}{k_B T_1} - \frac{\hbar\Omega_G}{k_B T_2}\right\} \tag{7}$$

obtained from $(1+n)/n \propto \exp(\hbar\Omega_G/k_b T)$. Here, $\hbar\Omega_G = 190$ meV is the energy of the G phonon and $n$ is the corresponding phonon occupation number, while $T_1 = T_0 + \Delta T_G$ and $T_2 = T_0 + 2\Delta T_G$ are the temperatures of the G phonon at 75 kW/cm$^2$ and 150 kW/cm$^2$, respectively. Assuming that the G phonon temperature is proportional to the laser excitation power, we obtain $\Delta T_G = 82$K by choosing $T_0 = 293$ K. Alternatively, we scale the Raman shifts of the G$^+$-contribution (see Fig. 7(b)) with respect to the thermal coefficient $d\Omega_{G^+}/dT$ = -0.02 cm$^{-1}$K$^{-1}$ [31] and obtain a temperature increase of $2\Delta T_G$ = 85K for the highest laser power density of 150 kW/cm$^2$, a factor of two smaller from that derived from the respective anti-Stokes/Stokes ratio.

## 4. Summary and conclusion

In order to evaluate the role of the substrate on the spectroscopic properties of a nanotube, we performed Raman excitation spectroscopy and PL excitation spectroscopy on a single, semiconducting single-walled carbon nanotube that is partially suspended across a trench and partially supported by a SiO$_2$-substrate. In the suspended CNT segment, the PL excitation spectrum is dominated by localized excitons and is red shifted by 40 meV with respect to the Raman excitation spectrum, which is determined by free excitons. In the supported CNT segment, the Raman excitation spectrum originates from localized excitons resulting in a red shift of 50 meV due to both exciton localization at defect sites and dielectric scaling of the free exciton binding energy. We then analyzed the factors that determine the observed variations of the Raman scattering intensities in suspended and supported segments of the same CNT and we quantified three contributions: (A) the broadening of the $E_{33}$-resonance, (B) the loss of oscillator strength due to exciton localization and (C) the modification of the intensity distributions of the focused laser excitation field. We associated the different line shapes of Raman transition bands observed in suspended and supported CNT segments with symmetry breaking and local distortions at defect sites in the presence of the dielectric substrate. We found that the phonon-mode-specific response to laser-irradiation differs significantly in the freely suspended and supported segments of the same CNT.

**Acknowledgements**

We thank Achim Hartschuh (LMU München, Germany) for a critical reading of the manuscript as well as Ado Jorio (UFMG and INMETRO, Belo Horizonte and Rio de Janeiro, Brazil), Stephanie Reich (FU Berlin, Germany) and Christian Thomsen (TU Berlin, Germany) for discussions.

**Figures and Figure Captions**

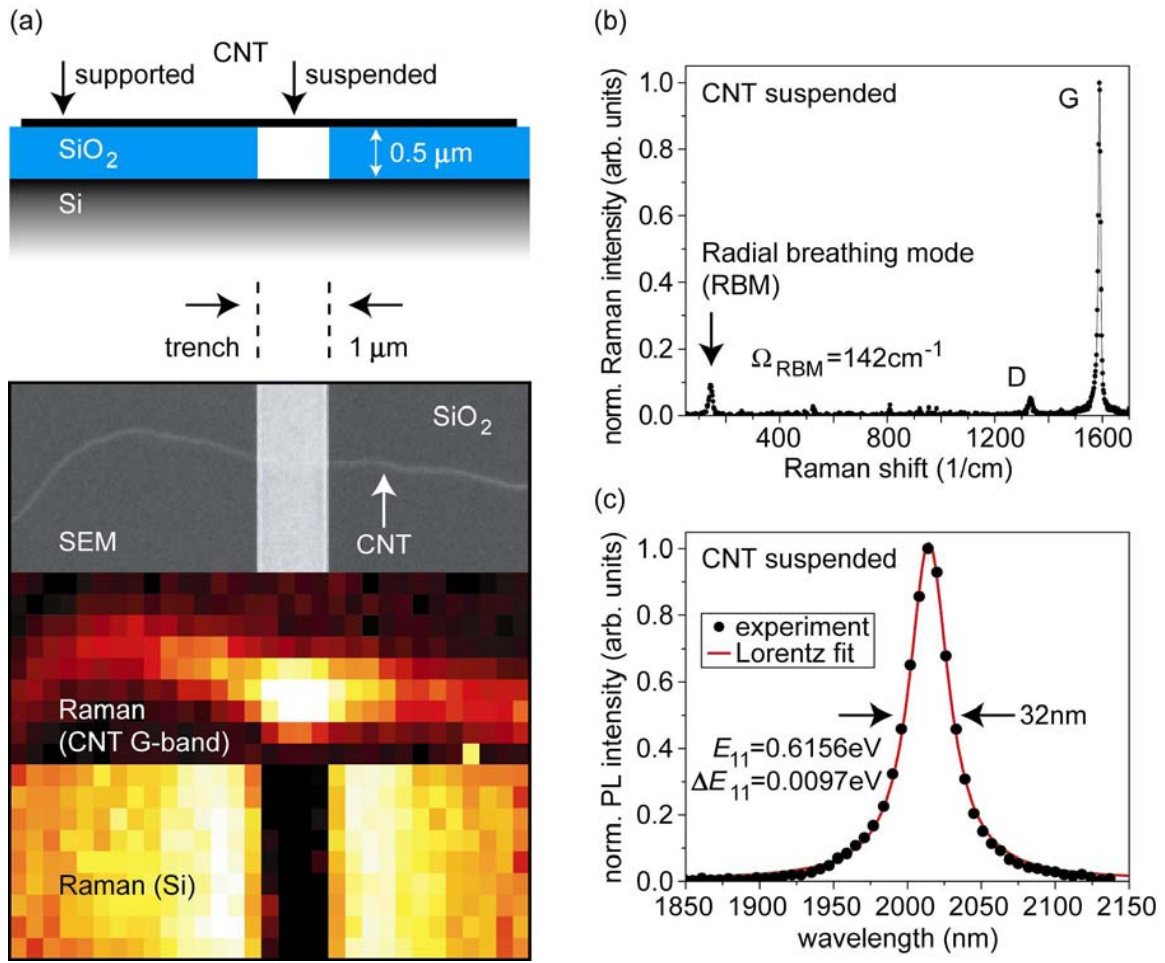

**Figure 1.** (a) Schematic cross section (side view), scanning electron microscope (SEM) and Raman ($\lambda_{laser}$=581.6nm) microscope images (top view) of the single-walled carbon nanotube (CNT) sample studied. (b) Raman spectrum ($\lambda_{laser}$=532 nm) and (c) photoluminescence (PL) spectrum ($\lambda_{laser}$=514.5nm) measured on the freely suspended part of the single, semiconducting CNT imaged in (a).

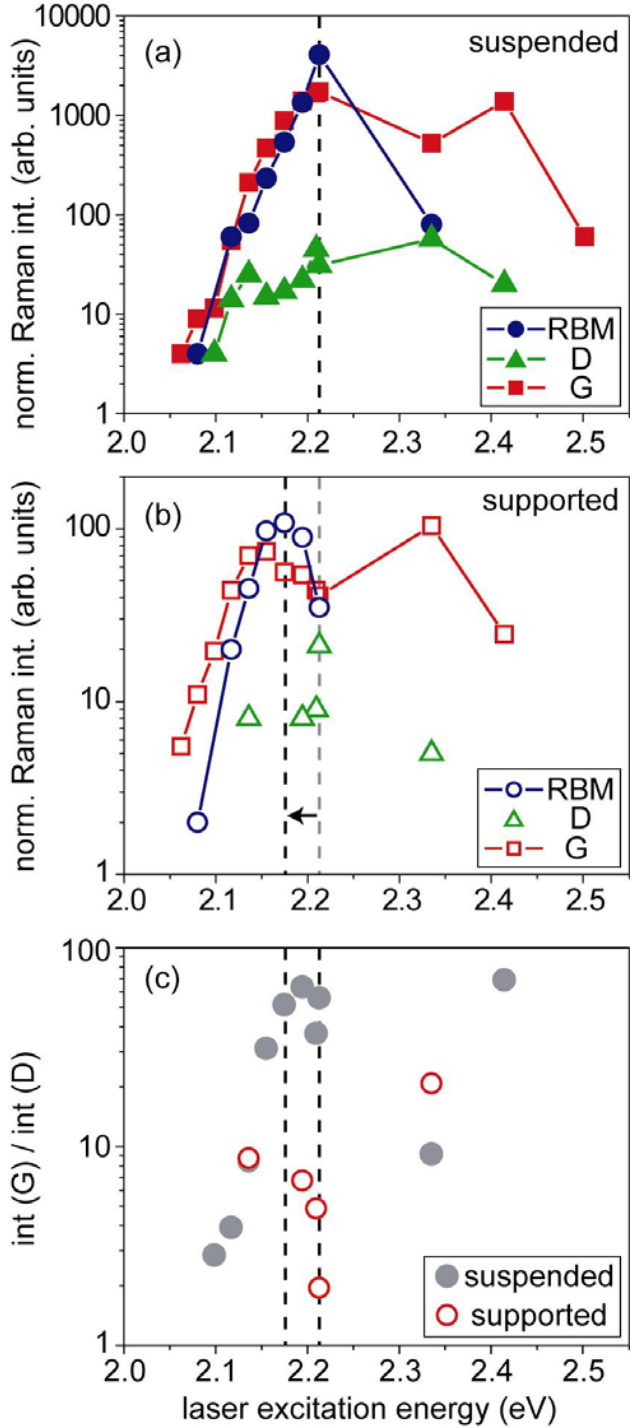

**Figure 2.** Resonance-Raman excitation spectra of the radial breathing mode (RBM), the D band and the G band measured on (a) the suspended and (b) the supported segment of the same, semiconducting CNT. The respective $E_{33}$-levels of the CNT are indicated by the vertical dashed lines. (c) Ratio of the integrated Raman intensities of the G band and the D band plotted as a function of the laser excitation energy.

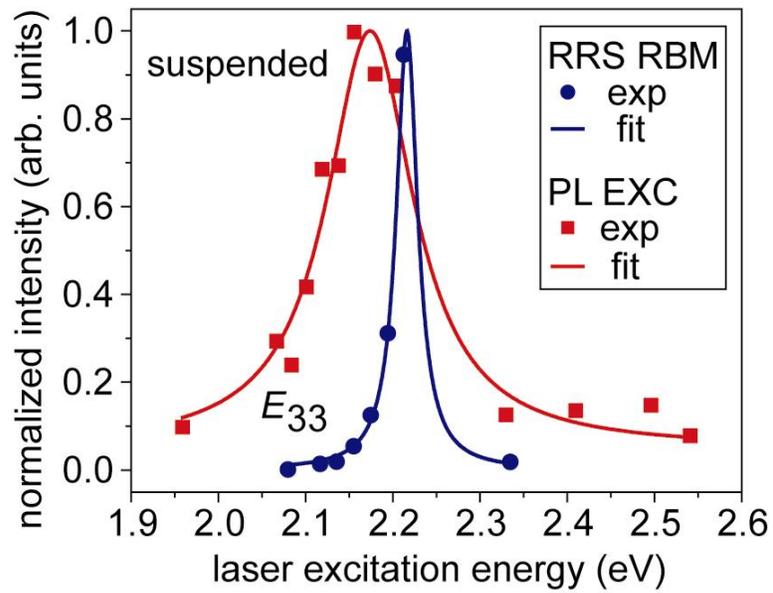

**Figure 3.** Resonance-Raman excitation spectrum of the integrated radial breathing mode intensity (RRS RBM) and excitation spectrum of the integrated $E_{11}$-photoluminescence intensity (PL EXC) reveal the $E_{33}$-exciton in the suspended CNT segment. The experimental spectra have been fitted by single Lorentzian line shape functions. The spectral peak positions of the two spectra are displaced by 40 meV.

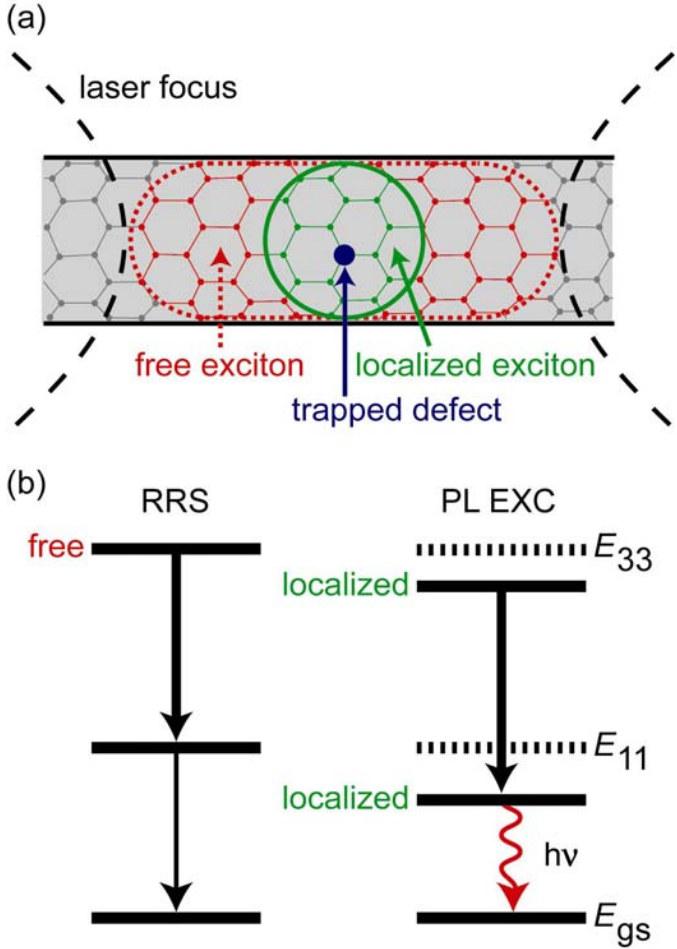

**Figure 4.** (a) Schematic indicating the localization of an optically excited exciton at a trapped defect site along the CNT. (b) The energy level diagram indicates the contributions of localized and free excitons in the two measured excitation spectra shown in Fig. (3). While the absorption by free $E_{33}$-excitons is probed in the resonance-Raman excitation spectrum (RRS), the absorption by localized $E_{33}$-excitons (energetically lower by 40 meV) is probed in the photoluminescence excitation spectrum (PL EXC).

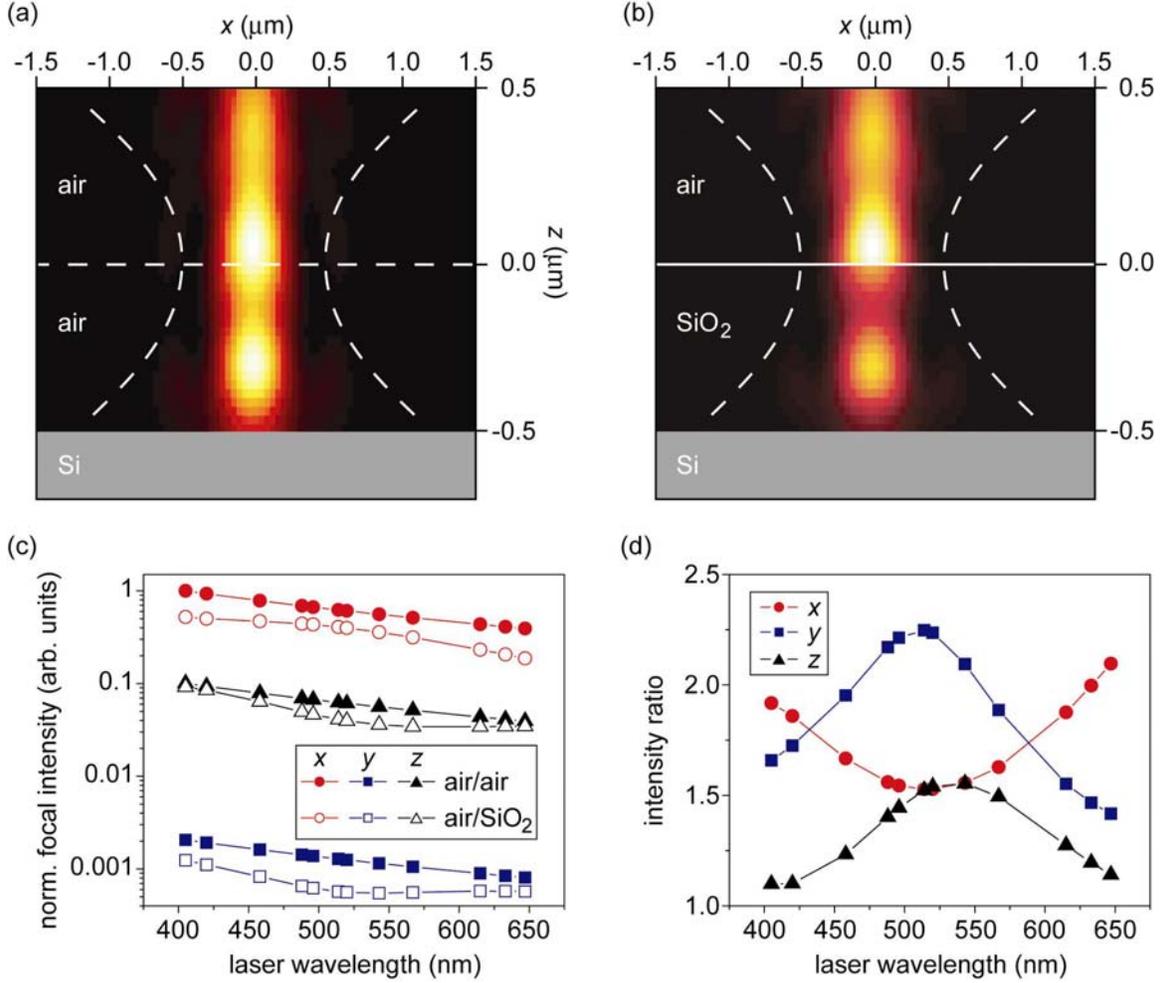

**Figure 5.** Calculated intensity distributions of the focused laser excitation field (Hermite-Gaussian TEM$_{00}$-mode, polarized in *x*-direction) at the position of the (a) suspended and (b) supported segment of the CNT located at *z*=0. The laser wavelength is $\lambda_{laser}$ = 567 nm. (c) Comparison of the focal intensities associated with each individual laser field component at the (*x*=0, *z*=0)-position in (a) and (b) as a function of the laser wavelength. The corresponding focal intensity ratios for each component shown in (d) display the enhancement factors of the laser excitation at the position of the suspended CNT segment.

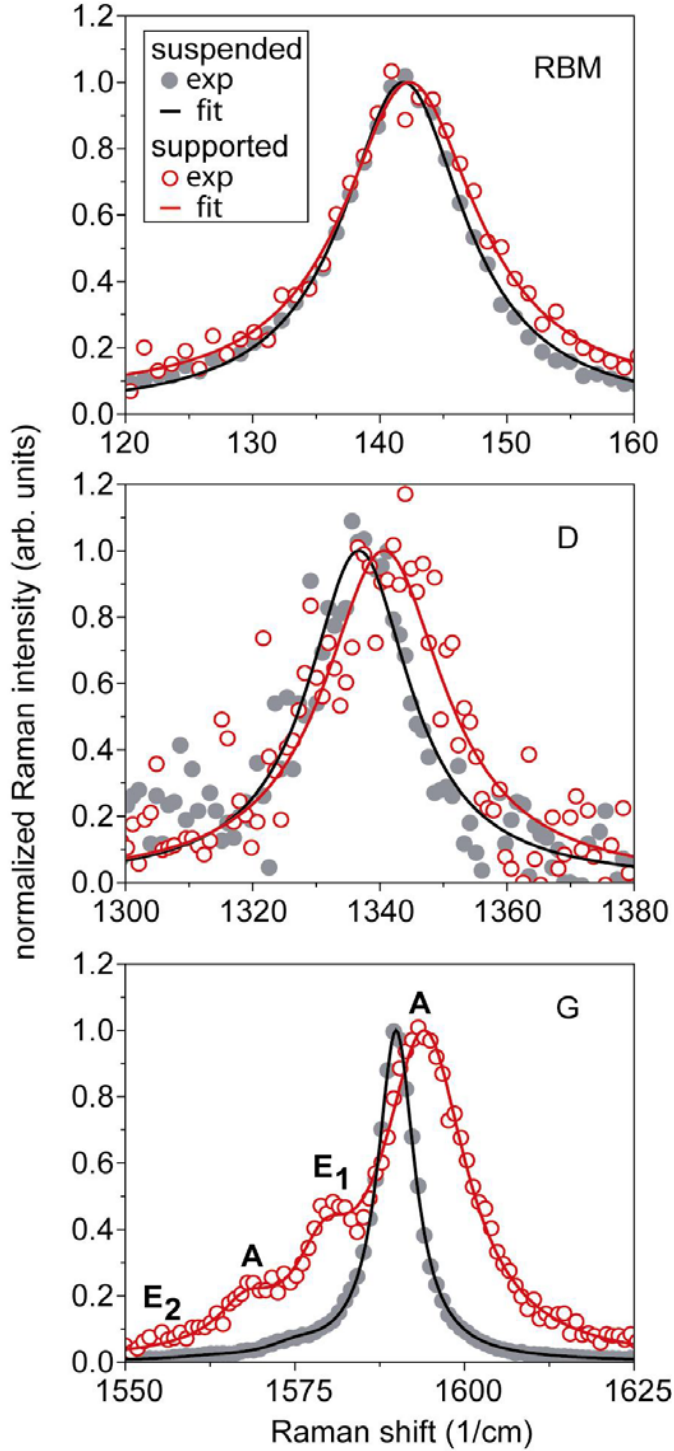

**Figure 6.** High-resolution Raman spectra of (top to bottom) the radial breathing mode (RBM), the D band and the G band measured on suspended and supported segments, respectively, of the same, semiconducting CNT and fitted to (multiple) Lorentzian line shape functions. The excitation laser ($\lambda_{laser}$ = 567nm) is tuned on resonance with the $E_{33}$-level in the supported CNT segment (~2.19 eV, see Fig.2). The assignment of phonon modes in the G-band spectrum is discussed in the text.

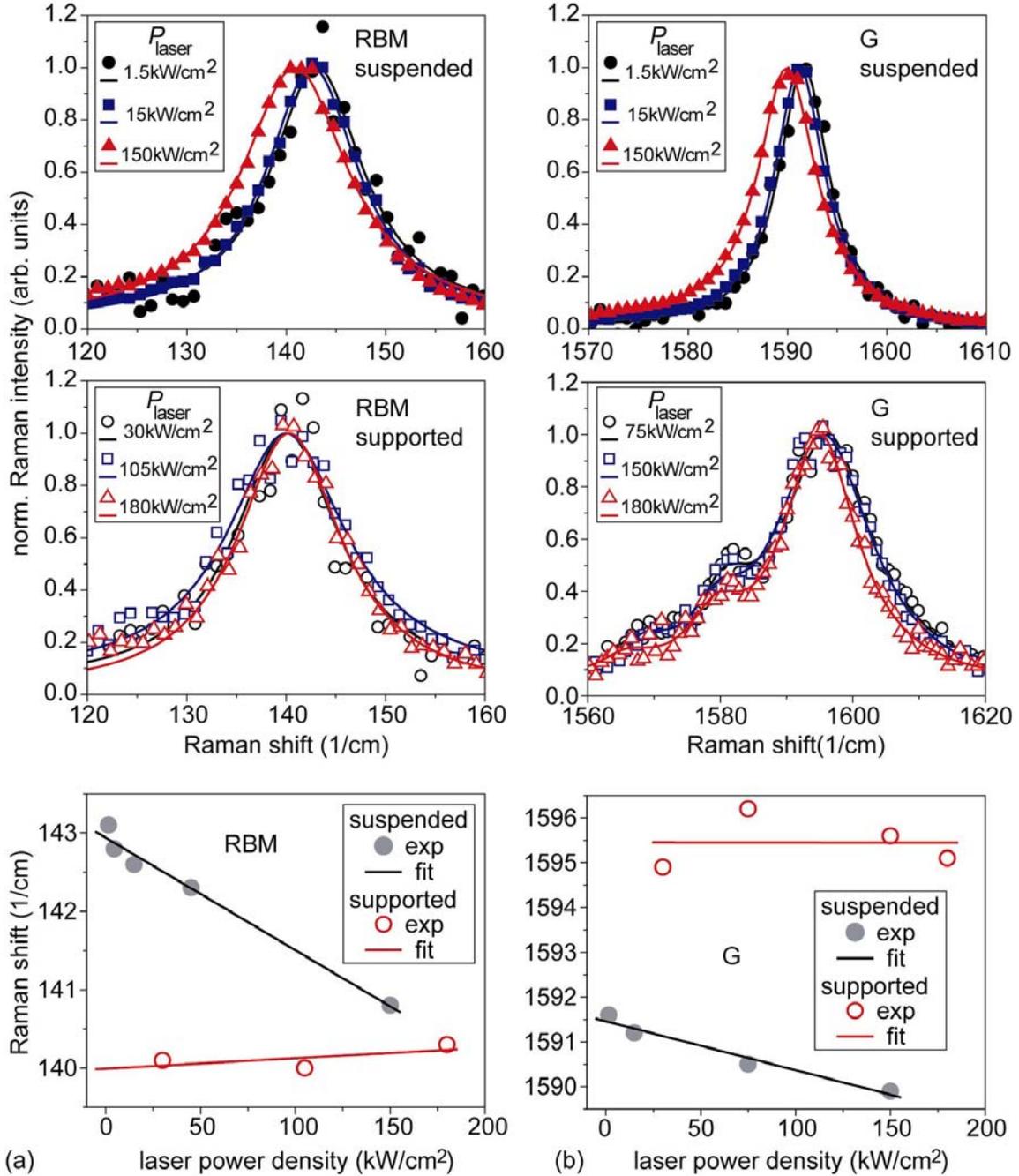

**Figure 7.** (a) Radial breathing mode (RBM) and (b) G band Raman spectra (symbols) measured on the suspended or the supported part, respectively, of the same semiconducting CNT at different laser power levels. Each spectrum is fitted to (multiple) Lorentzian line shape functions (lines). In the lower panels of (a) and (b), the extracted spectral peak positions are plotted as a function of the laser power density together with linear fits to the data. The laser excitation wavelength is $\lambda_{laser} = 567$ nm.

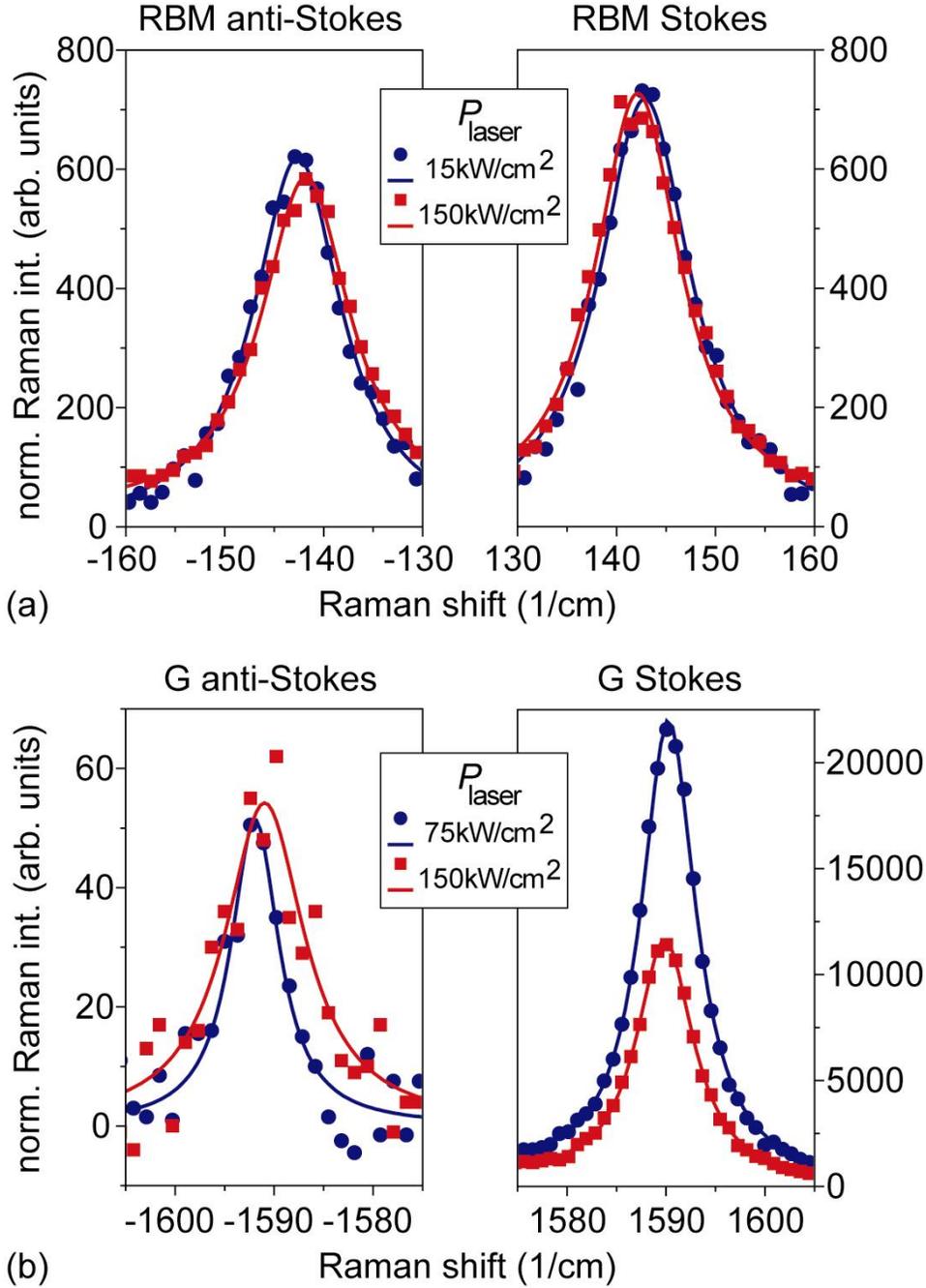

**Figure 8.** Stokes and anti-Stokes spectra of (a) the radial breathing mode (RBM) and (b) the G band, each measured on the suspended part of the same, semiconducting CNT for two representative laser excitation powers (symbols). For comparison, all spectra have been normalized with respect to laser excitation power density and data acquisition time. The Raman bands are fitted individually based on Lorentzian line shape functions (lines). The laser excitation wavelength is $\lambda_{exc}=$ 566nm.